\documentclass[superscriptaddress, reprint, amsmath, amssymb, aps, pra, floatfix]{revtex4-2}

\usepackage{mathptmx}                  % Times for math (legacy; ok for quick template)
\usepackage{xcolor}
\definecolor{dgreen}{rgb}{0.0,0.6,0.0}
\definecolor{pink}{rgb}{1,0,0.9}
\usepackage{physics}
\usepackage{graphicx}
\usepackage{hyperref}
\hypersetup{colorlinks=true, linkcolor=blue, filecolor=magenta, urlcolor=magenta}
\usepackage{mathrsfs}
\usepackage{orcidlink}
\usepackage{comment}
\usepackage{multirow}
\usepackage{changepage}
\begin{document}
\raggedbottom
\title{Digital Twin Modeling of Quantum Dynamical Systems: Dissipative Quantum Reservoir Computing}

\author{Abhijit Sen\,\orcidlink{0000-0003-2783-1763}}
\email{asen1@tulane.edu}
\affiliation{Department of Physics and Engineering Physics, Tulane University, New Orleans, Louisiana 70118, USA}

\author{Bikram Keshari Parida\,\orcidlink{0000-0003-1204-357X}}
\email{parida.bikram90.bkp@gmail.com}
\affiliation{Department of Physics and Engineering Physics, Tulane University, New Orleans, Louisiana 70118, USA}
\affiliation{Department of Data Science, Dongduk Women's University, Seoul, South Korea}

\author{Shital Chauhan\,\orcidlink{0000-0002-1817-7379}}
\email{shital9chauhan@gmail.com}
\affiliation{SSJDWSSSS Government Post Graduate College, Ranikhet 263645, India}

\author{Mahima Arya\,\orcidlink{0000-0002-1847-9705}}
\email{aryamahima@gmail.com}
\affiliation{Department of Physics and Engineering Physics, Tulane University, New Orleans, Louisiana 70118, USA}

\author{Denys I. Bondar\,\orcidlink{0000-0002-3626-4804}}
\email{dbondar@tulane.edu}
\affiliation{Department of Physics and Engineering Physics, Tulane University, New Orleans, Louisiana 70118, USA}

\date{\today}

% ------------------------- Example Abstract --------------------------
\begin{abstract}

Modeling the response of driven many-body quantum systems from input--output data is difficult: the dynamics are nonlinear, history dependent, and expensive to simulate as system size grows. A paradigmatic case is High-Harmonic Generation~(HHG), where a strong field drives a medium to emit radiation that is highly sensitive to the drive and encodes long-range temporal correlations. We introduce a dissipative quantum reservoir computing~(DQRC) framework that builds a digital twin of such a system, learning its input--output map directly from data while the reservoir---itself a small open quantum system---stays fixed and only a classical readout is trained. We show that a minimal single-qubit reservoir reproduces the HHG response of a substantially larger Ising spin chain, and on a representative benchmark matches and on several metrics surpasses previously reported temporal convolutional and Kolmogorov--Arnold-network models, while using a simpler, physically realizable system. A single fixed reservoir further generalizes across a broad range of drives, indicating that it learns a shared physical response structure rather than memorizing trajectories. These results establish dissipative quantum reservoirs as compact, physically grounded digital twins for nonlinear, memory-dependent quantum dynamics.

Code is available at \href{https://github.com/AI-and-Quantum-Computing/DQuRC}{https://github.com/AI-and-Quantum-Computing/DQuRC}.
\end{abstract}

\maketitle
\section{Introduction}

Much of physics can be cast as a time-series analysis problem. From a sequence of observed measurements, one infers the underlying dynamical law and uses it to forecast future evolution. A physically meaningful model must therefore possess predictive power. It should yield accurate estimators of observables at later times, given the present state and past history~\cite{Bondar2012}. Its strength lies not only in interpolating an observed series, but in extrapolating beyond it. Machine learning~(ML) approaches the same task from data rather than from a closed-form law. This makes it effective for forecasting nonlinear, high-dimensional dynamics. Conventional ML models, however, often reproduce trajectories through statistical correlation rather than physical law. As a result, they fail when forecasting beyond the training series. We observed this limitation directly in our purely black-box treatment of optical input--output dynamics, which used a generative temporal convolutional model~\cite{Sen2025TCN}. To overcome it, our subsequent work~\cite{Sen2025KAN} embedded physical structure directly into the learning architecture through Ehrenfest constraints~\cite{Ehrenfest1927, Bondar2012}. This yielded a significant improvement over the black-box approach~\cite{Sen2025KAN}. 

In this work we pursue a complementary route that further improves upon this physics-informed model. The idea is to combine ML with quantum reservoir computing~\cite{Fujii2017,Nakajima2019} as a trained surrogate. Given sufficient input--output data, this surrogate learns an effective representation of a complex system. It then predicts the system's observable response without solving the full microscopic dynamics, yielding a digital twin of the underlying system. This  viewpoint is especially useful for driven many-body quantum systems. Their input--output response is nonlinear, history dependent, and rapidly expensive to simulate as system size grows.

Among driven quantum phenomena, few are as demanding or as experimentally significant as the nonlinear optical response of laser-driven systems, exemplified by High-Harmonic Generation~(HHG), where a strong laser pulse drives a quantum medium to emit coherent radiation at integer multiples of the fundamental frequency. HHG underpins attosecond spectroscopy~\cite{Krausz2009} and, since its first observation in bulk crystalline solids~\cite{Ghimire2010}, has enabled precision measurement of the electronic band structures~\cite{Luu2015}, inspired proposals for single-atom computing~\cite{McCaul2023}, facilitated characterization of quantum mixtures~\cite{Magann2022}, and, in spin systems, provided a probe of spin dynamics and a route toward compact THz sources~\cite{Takayoshi2019,Ikeda2019,Malla2023}. Its output is exquisitely sensitive to the driving field, encodes long-range temporal correlations, and spans several orders of magnitude in intensity, making it a stringent test bed for any data-driven surrogate. ML has accordingly been applied across HHG tasks, including surrogate spectra and inverse parameter retrieval~\cite{Lytova2023}, microscopy~\cite{Shen2023}, solid-state spectroscopy~\cite{Klimkin2023}, plasma-driven feature prediction~\cite{Mihailescu2016}, spectral forecasting~\cite{Yan2022}, full TDSE--Maxwell macroscopic dynamics and structured-field simulation~\cite{Serrano2023,PablosMarin2023}, as well as ultrashort-pulse reconstruction~\cite{Zahavy2018,Brunner2022} and photoelectron denoising~\cite{KumarGiri2020}.

\begin{figure}[ht]
    \centering
    \includegraphics[width=0.85\linewidth]{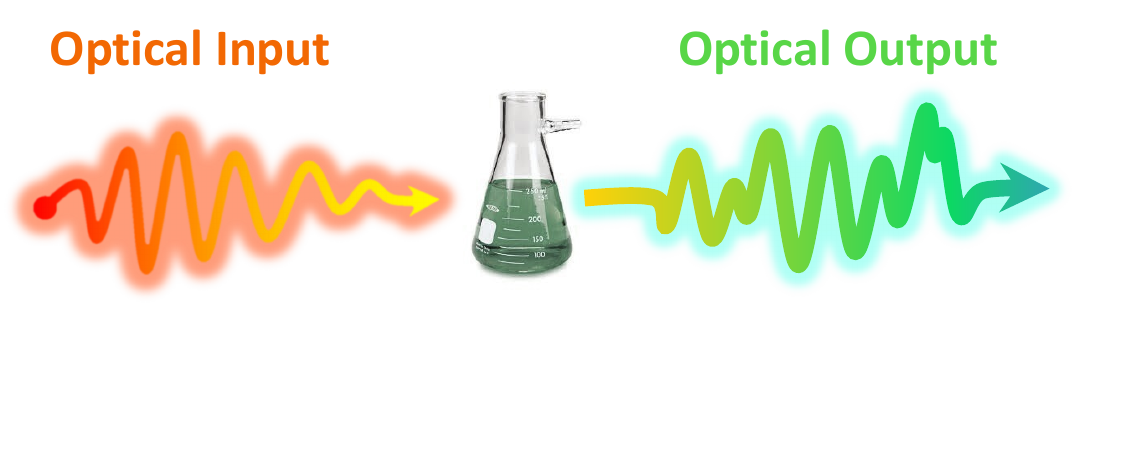}
    \includegraphics[width=\linewidth]{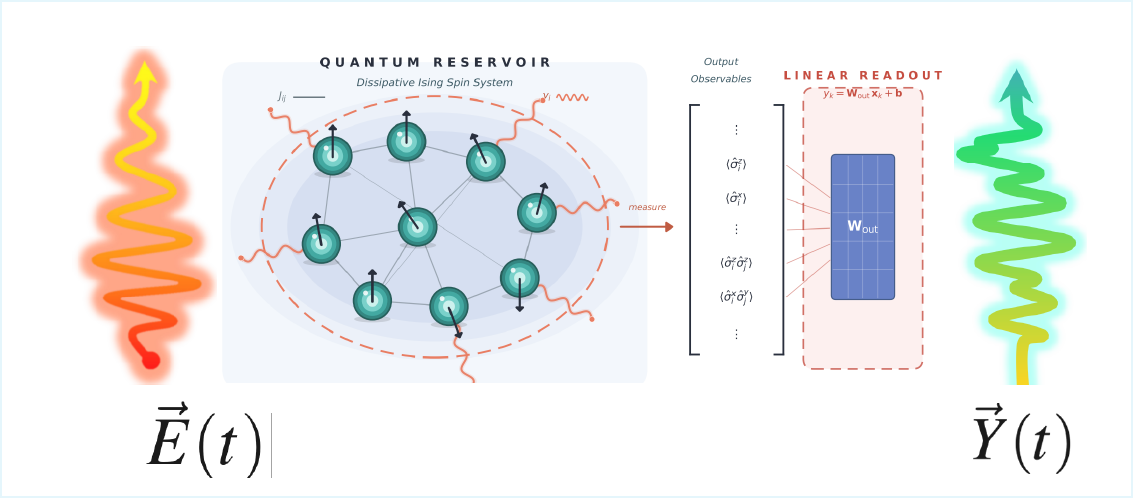}
    \caption{Optics as an input--output problem. An input field $\mathbf{E}(t)$ (red) interacts with a medium (beaker) to produce an output field $\mathbf{Y}(t)$ (green). A quantum Reservoir with trainable readout, the readout layer trained on $(\mathbf{E}(t), \mathbf{Y}(t))$ pairs, serves as a surrogate model of the medium, enabling prediction of $\mathbf{Y}(t)$ for previously unseen inputs $\mathbf{E}(t)$.}
    \label{figkan}
\end{figure}

We treat optics as an input--output problem, as depicted in Fig.~\ref{figkan}: an external field $\vec{E}(t)$ interacts with a physical medium (the quantum reservoir), which processes it through its intrinsic nonlinear, high-dimensional quantum dynamics to produce an output field $\vec{Y}(t)$. A trainable linear readout, optimized on recorded $(\vec{E}(t),\,\vec{Y}(t))$ pairs, turns the system into a data-driven surrogate that predicts $\vec{Y}(t)$ for previously unseen inputs without any explicit microscopic model. This is precisely the paradigm of Quantum Reservoir Computing~(QRC), which embeds temporal information into the Hilbert space of a fixed quantum many-body reservoir and extracts predictions through a simple linear readout~\cite{Fujii2017,Nakajima2019}.

Most prior QRC architectures rely on idealized closed unitary evolution, whereas optical and superconducting platforms are inherently open, subject to dissipation and decoherence. While such non-unitary effects have been studied through natural dissipation and measurement back-action, dissipation is usually treated as an intrinsic property of the platform and analyzed under fixed dynamics~\cite{Suzuki2022}. Here we instead treat dissipation as an explicitly engineered component of the reservoir, controlled through the Lindblad operators and rates, so that the memory kernel and dynamical contraction can be tuned to match the structure of the input signal. We govern the reservoir by the Lindblad master equation~(LME)~\cite{Lindblad1976,Gorini1976}, which guarantees completely positive and trace-preserving evolution and ties the reservoir's internal features to measurable observables without the externally imposed constraint penalties used in physics-informed learning. Engineered decoherence supplies a tunable fading-memory kernel that regulates information flow and suppresses chaotic divergence, so the resulting dissipative QRC~(DQRC) acts as a high-fidelity quantum transducer that maps a continuous drive to the evolution of a system observable. While HHG provides the physical motivation, we do not model microscopic HHG processes directly, but consider a minimal driven open quantum system that captures their essential structure as a temporal transduction problem.

It is important to clarify the scope of this study. We deliberately focus on the minimal one-body setting ($N=1$), in which neither entanglement nor exponential Hilbert-space scaling is present. The objective is therefore not to demonstrate quantum advantage, but to isolate driven dissipative dynamics as a minimal computational primitive for temporal processing, with memory and nonlinearity arising directly from the open-system evolution and stability following from the contractive nature of the Lindblad dissipator. This minimalism stands in sharp contrast to our previous physics-informed approach~\cite{Sen2025KAN}, where the same task was addressed with high-dimensional Kolmogorov--Arnold networks comprising several hundred-unit hidden layers and a correspondingly large number of trainable parameters; here the dynamical substrate is a single fixed qubit and only a lightweight classical readout is trained. Any predictive capability observed in this regime should therefore be attributed to the structure of the driven dissipative dynamics rather than to Hilbert-space scaling, entanglement, or the expressive capacity of a large trained network.

The remainder of this paper is organized as follows. Section~\ref{sec:QRC} introduces the QRC framework, first conceptually and then with full technical detail, including the Lindblad dynamics, readout architecture, and training procedure. Section~\ref{sec:problem_formulation} formulates the input--output prediction problem for driven Ising spin chains and describes the datasets employed. Section~\ref{sec:results} presents numerical results, and Section~\ref{sec:conclusion} concludes with a discussion of broader implications and future directions.

\section{Quantum Reservoir Computing}
\label{sec:QRC}

\begin{figure*}[t]
    \centering
    \includegraphics[width=0.90\linewidth]{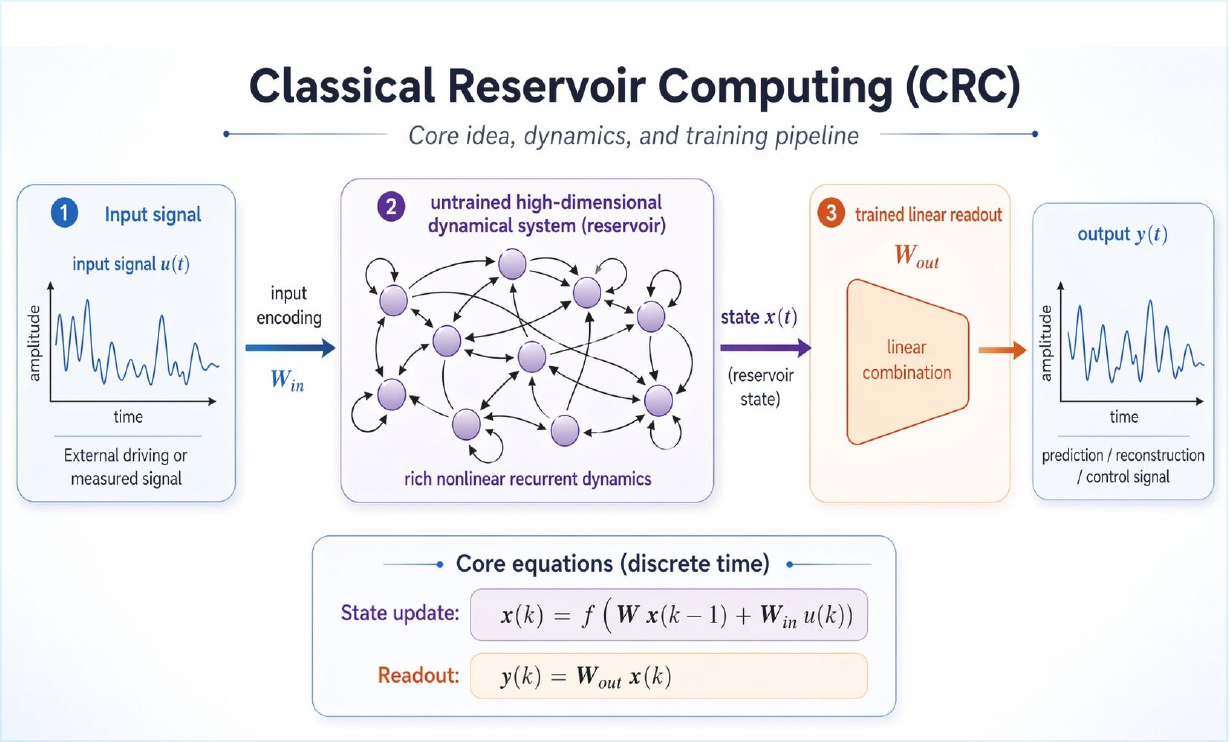}
    \caption{Schematic overview of Classical Reservoir Computing (CRC), illustrating the separation between a fixed nonlinear dynamical system (the reservoir) and a trainable linear readout. The input signal $u(t)$ is discretized and injected via the input weight matrix $W_{\mathrm{in}}$. The reservoir state $x(k)$ evolves according to $x(k) = f\!\left(W\,x(k-1) + W_{\mathrm{in}}\,u(k)\right)$, and the output $y(k)$ is obtained through a trained linear readout $y(k) = W_{\mathrm{out}}\,x(k)$, where only $W_{\mathrm{out}}$ is optimized during training.}
    \label{figkan1}
\end{figure*}

Reservoir computing (RC) provides a framework for temporal and sequential data, rooted in recurrent neural network (RNN) frameworks~\cite{Jaeger2001, Maass2002}. Its central idea is to delegate the computational burden to a high-dimensional dynamical system, the \textit{reservoir}, which is driven by the input signal but left entirely untrained. Learning is confined to a simple linear readout layer, trained via linear regression, while the reservoir's internal recurrent connections remain fixed. The nonlinear dynamics of the reservoir serve as an implicit feature map, so training reduces to a linear regression problem on the readout weights and the recurrent connections are never touched. 

Formally, at each discrete time step $k$, the reservoir state $\mathbf{x}(k) \in \mathbb{R}^N$ evolves as
\begin{equation}
    \mathbf{x}(k) = f\bigl(W\,\mathbf{x}(k-1) + W_{\mathrm{in}}\,\mathbf{u}(k)\bigr),
\end{equation}
and the output is obtained via
\begin{equation}
    \hat{y}(k) = W_{\mathrm{out}}\,\mathbf{x}(k),
\end{equation}
where $W$ and $W_{\mathrm{in}}$ are fixed random matrices and only $W_{\mathrm{out}}$ is trained. The reservoir acts as a nonlinear projection of the input into a high-dimensional state space, rendering complex temporal features more separable. Its recurrent dynamics naturally encode input history, endowing the system with fading memory, whereby the influence of past inputs on the current reservoir state decays over time. A foundational requirement for stable operation is the \textit{echo state property}~\cite{Jaeger2001,Maass2002}: the reservoir state must be uniquely determined by the input history alone, independently of the initial conditions. This ensures that transient dynamics die out and the reservoir settles onto a well-defined input-driven trajectory. The rate at which memory fades must be matched to the characteristic timescales of the driving signal (too short a memory loses relevant history, while too long a memory conflates distinct inputs).
The fading-memory property of reservoir systems has been formalized in terms of memory capacity, which quantifies the ability of the reservoir to reconstruct past inputs from its current state. In classical reservoir computing, this has been analyzed in detail by Dambre \emph{et al.}~\cite{Dambre2012}, who showed that memory and nonlinear computational capacity are fundamentally constrained by the dimensionality and dynamics of the reservoir. In the present dissipative quantum setting, the Lindblad dynamics induce a contractive evolution that enforces fading memory through exponential decay of past inputs, suggesting a direct correspondence between the dissipation rate and the effective memory depth of the reservoir. While a full capacity analysis is beyond the scope of this work, the observed dependence of prediction accuracy on driving frequency (Sec.~\ref{sec:results}) is consistent with this framework, with lower-frequency inputs requiring longer memory horizons than those supported by the fixed dissipation rate.

This property is particularly advantageous for physical implementations, where only the readout requires training while the internal dynamics are left untouched. Such \textit{physical reservoir computers} have been demonstrated across a wide range of substrates, including photonic reservoirs, spintronic oscillators, compliant mechanical structures, and analog electronic circuits~\cite{Tanaka2019}, and even a single atom~\cite{McCaul2023}, supporting the view that any sufficiently nonlinear, high-dimensional dynamical system can serve as a reservoir. Despite these successes, classical reservoirs are ultimately constrained by the polynomial scaling of their effective feature-space dimension with system size, motivating the extension to quantum systems, whose Hilbert spaces grow exponentially.

Quantum reservoir computing (QRC) extends this paradigm by replacing the classical dynamical system with a quantum many-body system, exploiting quantum effects, quantum superposition, and entanglement to access a Hilbert space that grows exponentially with qubit number~\cite{Fujii2017,Nakajima2019}, a resource unavailable to any classical reservoir of comparable physical size. A system of $N$ qubits is described by a $2^N \times 2^N$ density matrix $\rho$, defined on an exponentially large Hilbert space, which defines the reservoir state. The structural logic of RC is preserved: the quantum reservoir dynamics remain fixed and untrained, while only the linear readout layer is optimized. Observables $a_i(t) = \mathrm{Tr}[\rho(t)\, A_i]$ form the output feature vector $\mathbf{x}(t)$, which is fed to the readout layer. While the underlying Hilbert space grows exponentially with system size, it is accessed through a finite set of observables, rendering the effective computational advantage more nuanced than the raw dimensional scaling suggests.

\begin{figure*}[t]
    \centering
    \includegraphics[width=0.90\linewidth]{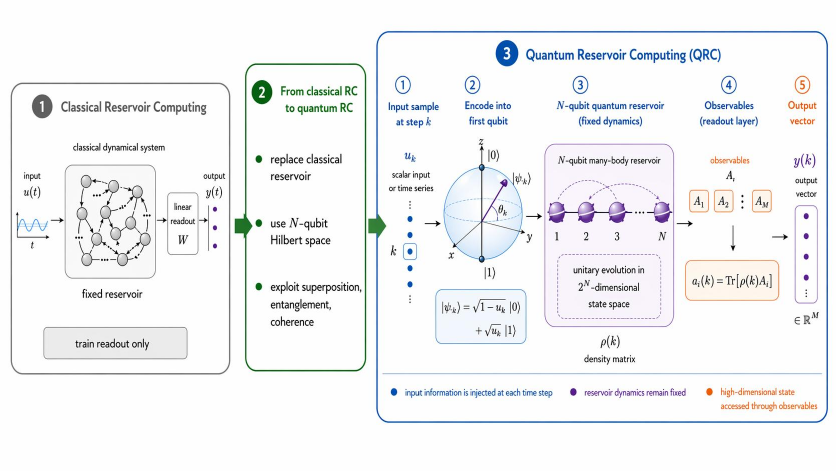}
    \caption{Schematic illustration of the transition from classical reservoir computing (CRC) to quantum reservoir computing (QRC). (Left) A classical dynamical system with a fixed reservoir and trained readout layer. (Center) Replacement of the classical reservoir with an $N$-qubit quantum system. (Right) Full QRC architecture: input is encoded at each time step into the first qubit, the reservoir evolves unitarily in the $2^N$-dimensional Hilbert space, and observables are extracted via a classical readout layer to produce the output vector.}
    \label{figkan2}
\end{figure*}

Input information is encoded by replacing the state of the first qubit at each discrete time step $k\tau$ with the pure state
\begin{equation}
    \rho_{u_k} = \ket{\psi_{u_k}}\bra{\psi_{u_k}}, \qquad
    \ket{\psi_{u_k}} = \sqrt{1 - u_k}\,\ket{0} + \sqrt{u_k}\,\ket{1},
\end{equation}
where $u_k \in [0,1]$ is the input value at step $k$. This encoding maps the scalar input to a point on a Bloch sphere meridian, with $u_k$ controlling the excitation probability, thereby preserving a direct physical interpretation of the input signal. The full density matrix then evolves unitarily under the system Hamiltonian, inducing a nonlinear trajectory in the large Hilbert space.

Since the full $2^N$-dimensional quantum state is experimentally inaccessible, the reservoir is probed through a finite set of local observables. To enrich the feature representation without increasing the number of qubits, \textit{temporal multiplexing} is employed: the output signals are sampled at $V$ intermediate times within each input interval, increasing the number of effective reservoir nodes from $N$ to $NV$. This allows the exponentially large Hilbert space to be probed using only a polynomial number of measurement signals~\cite{Fujii2017}, making the scheme experimentally tractable.

QRC may be formulated in discrete or continuous time; the latter is especially natural for physically driven systems, where the reservoir dynamics are governed by quantum master equations within the theory of open quantum systems. In this setting, interactions with the environment introduce dissipation and decoherence which, while conventionally regarded as detrimental to quantum computation, have been shown to markedly improve the dynamical stability, noise resilience, and memory characteristics of quantum reservoirs when deliberately engineered~\cite{Domingo2023,Monzani2024}. Engineered dissipation can itself enrich the reservoir response~\cite{Sannia2024}, reframing dissipation as a computational asset rather than a source of error.

As a controlled and analytically tractable testbed, we consider a driven Lindblad system under monochromatic sinusoidal driving, capturing the nonlinear response and memory effects that characterize driven–dissipative systems in the HHG regime.

High harmonic generation (HHG) is a canonical example of a nonperturbative, ultrafast process in which emitted high-frequency radiation encodes both the driving field and the underlying quantum dynamics of the medium~\cite{Lewenstein1994, Krausz2009}. From a machine-learning perspective, HHG constitutes a demanding temporal transduction problem: a continuous, broadband driving waveform must be faithfully mapped to a highly structured, history-dependent spectral output. The intrinsic high-dimensionality and temporal memory of quantum reservoirs make QRC a compelling and physically transparent framework for capturing such field-driven, nonlinear responses without incurring the prohibitive cost of first-principles simulations.

Among open-system QRC architectures, dissipative quantum reservoirs (DQRs) are well suited to continuous temporal signal processing. Coherent reservoirs are prone to uncontrolled growth of quantum correlations and are sensitive to initial conditions; in contrast, DQRs naturally relax to input-dependent steady states through coupling to the environment. Dissipation draws the reservoir toward an input-driven trajectory regardless of where it started, naturally implementing the fading-memory condition without any architectural tuning. This motivates the use of dissipative quantum reservoirs as the central framework in the present work.

In the context of HHG, this fading memory ensures that the induced dipole moment at time $t$ is 
governed by the recent history of the laser pulse rather than the infinite past, in direct correspondence with the physical picture of ultrafast electron dynamics. Furthermore, controlled dissipation also sharpens the reservoir's nonlinear feature-mapping capability, enhancing separability in Hilbert space and improving performance in complex temporal learning tasks~\cite{Sannia2024}. DQRC is therefore not merely compatible with driven open-system physics; it is shaped by the same principles that govern the target response.

\section{Problem Formulation}
\label{sec:problem_formulation}
To probe reservoir capacity fairly, we vary the driving protocol in a controlled hierarchy of complexity. Under a
single-amplitude excitation, the dimensionality of the input is deliberately
minimal, so that intrinsic reservoir properties such as memory depth, sensitivity to
initial conditions, and the geometry of Hilbert-space trajectories can be
isolated from any confounding structure in the drive itself. A subsequent
multi-amplitude protocol then tests whether that capacity survives when the input
richness is substantially increased; capacity that degrades under richer inputs
reflects a fragile dependence on specific dynamical conditions rather than a
general computational resource. The two protocols are thus complementary: single-amplitude results define a baseline, and multi-amplitude performance is read against it.

\subsection{Physical system and dataset construction}

We consider a driven quantum system that defines a continuous-time input--output map $h(t) \rightarrow Y(t)$, from an externally applied control field to a measured observable. All datasets used for training and evaluation are generated from this single physical model by varying the driving field $h(t)$. The data-generating system is a transverse-field Ising chain of $N = 10$ spins with equally spaced site-dependent couplings $\left( J_z^{(i)} \right)_{i=1}^N = (0.784, \ldots, 0.800)$, whose total Hamiltonian reads
\begin{align}
H(t) &= H_0 + h(t)\,V,
\label{eq:Ht_full} \\
H_0 &= - \sum_{i=1}^{N} J_z^{(i)} \, \sigma_z^{(i)} \sigma_z^{(i+1)}, \label{eq:H0}\\
V &= - \sum_{i=1}^{N} \sigma_x^{(i)}.
\label{eq:V}
\end{align}
Here the  periodic boundary conditions $\sigma_z^{(N+1)} \equiv \sigma_z^{(1)}$ are used, so the chain forms a closed loop with $N$ interaction bonds. Monochromatic drivings $h(t) = F_0 \sin(\omega t)$ are used throughout. 

Note that this data-generating system is deliberately high-dimensional ($N=10$), whereas the reservoir itself operates in the minimal single-qubit setting [see Sec.~\ref{SecReservoir} below], so that any predictive capability must be attributed to the driven-dissipative dynamics rather than to Hilbert-space scaling or entanglement.

Time evolution is obtained by unitary Schr\"odinger evolution, solved numerically using the QuTiP library. The output observable is defined as
\begin{equation}
Y(t) = \langle \psi(t) | V | \psi(t) \rangle,
\end{equation}

yielding one trajectory per driving field $h(t)$.

The driving frequency $\omega$ is sampled over a dense grid spanning sub- and super-resonant regimes relative to the system energy gap. Each trajectory is evolved over a fixed time window corresponding to one period of the lowest driving frequency and discretized into $512$ time steps. 

Each sample therefore corresponds to one complete dynamical trajectory. The input and output signals are discretized on a uniform grid of length $T=512$,
\begin{equation}
h^{(i)}_k = h(k\Delta t), 
\qquad 
Y^{(i)}_k = Y(k\Delta t),
\qquad 
k=1,\ldots,T .
\end{equation}
Thus, one complete system run gives an input vector
\begin{equation}
\mathbf{h}^{(i)}
=
\left[
h^{(i)}_1,
h^{(i)}_2,
\ldots,
h^{(i)}_{512}
\right],
\end{equation}
and the corresponding output vector
\begin{equation}
\mathbf{Y}^{(i)}
=
\left[
Y^{(i)}_1,
Y^{(i)}_2,
\ldots,
Y^{(i)}_{512}
\right].
\end{equation}
Here, the superscript $i$ labels the $i$-th system run. Each pair
\begin{equation}
\mathbf{h}^{(i)} \longrightarrow \mathbf{Y}^{(i)}
\end{equation}
therefore represents one complete input--output trajectory of the physical system.

The full dataset is arranged as a collection of such paired sequences:
\begin{equation}
\mathscr{D}(F_0)
=
\left\{
\mathbf{h}^{(i)}
\rightarrow
\mathbf{Y}^{(i)}
\right\}_{i=1}^{n},
\end{equation}
where $n$ is the total number of available system runs.

In the multi-amplitude setting, the dataset is constructed by evaluating $\mathscr{D}(F_0) $ across discrete amplitudes $F_0 \in \{1,2,\dots,20\}$, concatenating the resulting trajectory sets. For each amplitude, the full frequency grid is traversed, generating one trajectory per $(F_0, \omega)$ pair. A 80:20 partition of $\mathscr{D}(F_0) $ into training and testing data sets is used.

\subsection{Reservoir}\label{SecReservoir}

We employ a DQR  of $N$ qubits given by the Lindblad master equation
\begin{align}
  \frac{d\rho}{dt}
  &= -i\bigl[H_{\mathrm{res}}(t),\,\rho(t)\bigr]
  + \sum_{i=1}^N \gamma
  \Bigl(
    \sigma_i^{-}\,\rho\,\sigma_i^{+}
    - \tfrac{1}{2}
    \bigl\{\sigma_i^{+}\sigma_i^{-},\,\rho\bigr\}
  \Bigr),
  \label{eq:LME}\\
  & H_{\mathrm{res}}(t) = \left(\tfrac{1}{2} + g\,h(t)\right)\sum_{i=1}^{N}\sigma_x^{(i)},
\label{eq:Hres_general}
\end{align}
The data-generating system is deliberately chosen to be high-dimensional, while the reservoir itself operates in the minimal $N=1$ setting.

The target observable is the expectation value of the longitudinal spin
operator,
\begin{equation}
  y(t) = \sum_{i=1}^{N} \mathrm{Tr}\bigl[\sigma_z^{(i)}\,\rho(t)\bigr],
  \label{eq:target_obs}
\end{equation}
obtained by integrating the Lindblad master equation~\eqref{eq:LME} from
the ground state $\rho(0) = |0\rangle\langle 0|$ using the \textsc{QuTiP} (version 5.2.3)
\texttt{mesolve} solver~\cite{Johansson2012,Johansson2013}. This initialisation is
consistent with the dissipative dynamics, as the collapse operator drives
the system toward $|0\rangle$, ensuring that transient effects from the initial condition are minimal. The resulting dataset of input-output trajectories $\{h(t),\,y(t)\}$, parametrized by the driving amplitude $F_0$ and frequency $\omega$, defines the benchmark for evaluating transduction fidelity. 
\begin{figure*}
    \centering
    \includegraphics[width=0.99\linewidth]{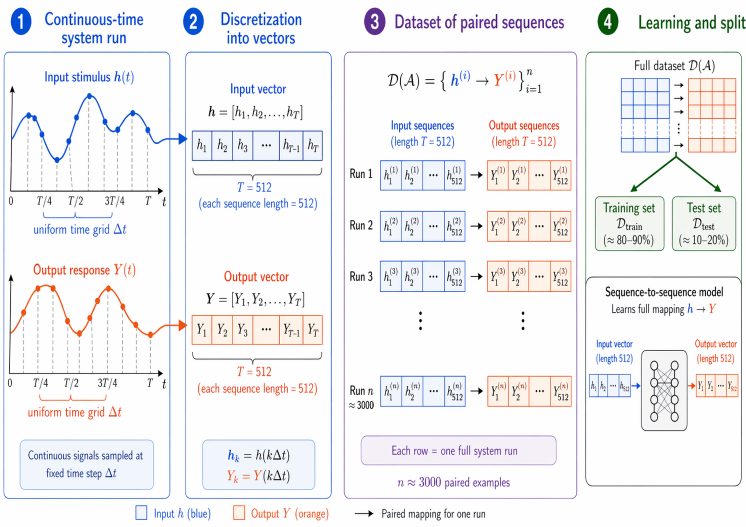}
    \caption{
Dataset construction and feeding strategy for input--output time-series learning with quantum reservoir computing. Continuous-time input stimuli $h(t)$ and output responses $Y(t)$ are first sampled on a uniform grid with time step $\Delta t$, producing fixed-length vectors of length $T=512$. Each row of the dataset corresponds to one complete system run, represented as a paired sequence $\mathbf{h}^{(i)} \rightarrow \mathbf{Y}^{(i)}$. The full dataset $\mathscr{D}(F_0)$ is split into training and test subsets. During learning, each input sequence is fed sequentially into the quantum reservoir, whose measured observables generate temporal reservoir features. A trainable readout then maps these features to the predicted output sequence.
}
    \label{fig:dataset1}
\end{figure*}

\subsection{Preprocessing and data partitioning}

Both the input signals and the ground-truth observables are independently
rescaled to $[0,1]$ via min-max normalization applied globally across all
trajectories,
\begin{equation}
  \tilde{u}_i = \frac{u_i - u_{\min}}{u_{\max} - u_{\min}},
  \label{eq:minmax}
\end{equation}
where $u_{\min}$ and $u_{\max}$ are computed over the full dataset. Global normalization is used rather than per-trajectory scaling in order to preserve amplitude-dependent variations across frequencies, which carry physically meaningful structure, while maintaining numerical stability during training.

\subsection{Reservoir embedding and readout}

Each normalized input signal $\tilde{h}(t)$ is passed through the dissipative
quantum reservoir. The reservoir input-gain parameter is fixed at $g=5.0$ throughout all simulations, and the
reservoir density matrix is evolved under Eq.~(\ref{eq:LME}) over
$T = 512$ time steps. At each step, the expectation value~\eqref{eq:target_obs} is recorded, yielding a temporal feature vector of dimension 512 per input trajectory; for a single-qubit reservoir ($N=1$ in Eq.~\eqref{eq:LME}), this corresponds to a feature tensor of shape  $(1 \times 512)$.

Principal component analysis (PCA) \cite{Jolliffe2002} is then applied to the ensemble of training features to reduce dimensionality and suppress noise from rapid transient fluctuations. The leading $n_{\mathrm{PCA}} = 15$ components are retained and the projection is fitted exclusively on the training set and applied without
refitting to the test set, preventing information leakage.

The resulting feature matrices are used to train a Kernel Ridge Regression (KRR) \cite{saunders1998kernel} readout with a radial basis
function (RBF) kernel,
\begin{equation}
  k(\mathbf{x},\mathbf{x}') =
  \exp\!\bigl(-\gamma_{\mathrm{rbf}}\,\|\mathbf{x}-\mathbf{x}'\|^{2}\bigr),
  \label{eq:rbf_kernel}
\end{equation}
Here, $\mathbf{x}, \mathbf{x}' \in \mathbb{R}^{15}$ denote the PCA-compressed feature vectors, which are compared pairwise during training and prediction. The RBF kernel implicitly maps these features into a reproducing kernel Hilbert space and enables nonlinear regression in closed form. The regularisation strength $\alpha$ and kernel bandwidth $\gamma_{\mathrm{rbf}}$ are selected by five-fold cross-validation over predefined grids, and the final model is refit on the full training set prior to evaluation on held-out test set.
This pipeline yields a nonlinear mapping from reservoir-generated dynamical features to the target observable. 

\subsection{Evaluation metrics}

Predictive performance is evaluated using mean squared error,
\begin{equation}
  \mathrm{MSE} = \frac{1}{N_{\mathrm{test}}\,T}
  \sum_{i=1}^{N_{\mathrm{test}}} \sum_{n=1}^{T}
  \bigl[\hat{y}_i(t_n) - y_i(t_n)\bigr]^{2},
  \label{eq:mse}
\end{equation}
which measures the average prediction error and the coefficient of
determination,
\begin{equation}
  R^{2} = 1 - \frac{%
    \sum_{i,n}\bigl[\hat{y}_i(t_n)-y_i(t_n)\bigr]^{2}}{%
    \sum_{i,n}\bigl[y_i(t_n)-\bar{y}\bigr]^{2}},
  \label{eq:r2}
\end{equation}
where $\bar{y}$ is the global mean of the ground-truth trajectories and
$R^{2}$ quantifies the fraction of variance explained by the model. In addition to aggregate metrics, performance is analyzed at the level of individual test trajectories by computing per-frequency $R^{2}$ scores as a function of the driving frequency $\omega$, providing a frequency-resolved view of transduction accuracy. This is particularly informative near resonance ($\omega \approx 1$) of the reservoir, where the system exhibits strongly nonlinear response and the prediction task is correspondingly most challenging. 

\section{Results and Discussion}
\label{sec:results}

The reservoir parameters are initialized from a fixed random seed, ensuring reproducibility across all reported results.

The dissipative quantum reservoir is evaluated across a sequence of driving amplitudes $F_0$, progressing from a weak-drive regime to a broad multi-amplitude setting. This hierarchy enables a controlled assessment of transduction accuracy as signal complexity increases: the single-amplitude case isolates intrinsic reservoir behavior, while the multi-amplitude regime probes generalization across a parametrized family of inputs.

\subsection{Weak-drive regime: $F_0 = 10$}

We first evaluate the quantum reservoir computing framework 
in the weak-drive regime, corresponding to $F_0 = 10$, 
using a minimal $N=1$ reservoir [Eq.~\eqref{eq:LME}]. This setting is directly 
comparable to dataset 2 of Ref.~\cite{Sen2025KAN}, which 
considers driving amplitude $A = 10$ (equivalent to $F_0=10$ in our notation) on the transverse-field 
Ising model.

Table~\ref{tab:f0_10} summarizes reconstruction performance 
for dataset sizes of $N_\omega = 200$ and $400$ trajectories. 
With $200$ samples, the reservoir achieves a mean $R^2 = 0.978$ 
across the test set. Increasing to $400$ samples yields 
near-perfect reconstruction with mean $R^2 = 0.9954$, with 
$97.5\%$ of test cases exceeding $R^2 > 0.90$ and $93.75\%$ 
exceeding $R^2 > 0.95$. The residual cases below threshold 
correspond predominantly to the lowest driving frequencies, 
where the output signal is most slowly varying and the 
reservoir requires greater temporal context to resolve 
fine structure --- a pattern consistent with observations 
in KAN-based approaches~\cite{Sen2025KAN}.

\begin{table}[h]
\centering
\caption{Reconstruction performance in the weak-drive regime 
($F_0 = 10$) for QRC with $N=1$ reservoir [Eq.~\eqref{eq:LME}], compared against 
KAN-based results from Ref.~\cite{Sen2025KAN} at equivalent 
amplitude ($A = 10$). Results report the percentage of test 
cases exceeding threshold values of $R^2 > 0.90$ and
$R^2 > 0.95$.}
\label{tab:f0_10}
\setlength{\tabcolsep}{2pt}
\begin{tabular}{llccc}
\hline\hline
Method & Architecture & $N_\omega$ & 
$R^2 > 0.90$ (\%) & $R^2 > 0.95$ (\%) \\
\hline
QRC& $N=1$ qubit        & 200 & 90& 75\\
QRC & $N=1$ qubit        & 400 & 97.5 & 93.75 \\
\hline
KAN& {[500,100,500]}     & 200 & 82.5 & 75.0  \\
KAN& {[500,400,400,500]} & 600 & 95.8 & 95.8  \\
\hline\hline
\end{tabular}
\end{table}

Notably, this level of accuracy is achieved by a single-qubit 
reservoir [$N=1$ in Eq.~\eqref{eq:LME}], in contrast to the KAN architecture of 
Ref.~\cite{Sen2025KAN}, which employs a $[500, 100, 500]$ 
network with $500$-dimensional input-output layers. The 
comparable reconstruction quality obtained here with a 
substantially simpler dynamical system underscores the 
transduction capacity of driven dissipative quantum dynamics, 
independently of Hilbert space scaling.

\subsection{Intermediate drive regime ($F_0 = 20$)}
With baseline performance benchmarked at $F_0$=10, we turn to a considerably more demanding dynamical regime by fixing the drive amplitude at $F_0 = 20$ and sweeping the driving frequency $\omega$ across both sub- and super-resonant bands. At this amplitude, the driven Ising chain enters a strongly nonlinear response regime, and the target trajectories exhibit considerably richer temporal structure than those observed at $F_0 = 10$. The dissipation rate is held fixed at $\gamma = 5.0$ throughout, so that performance variation reflects changes in drive complexity alone.

\begin{figure}
    \centering
    \includegraphics[width=\linewidth]{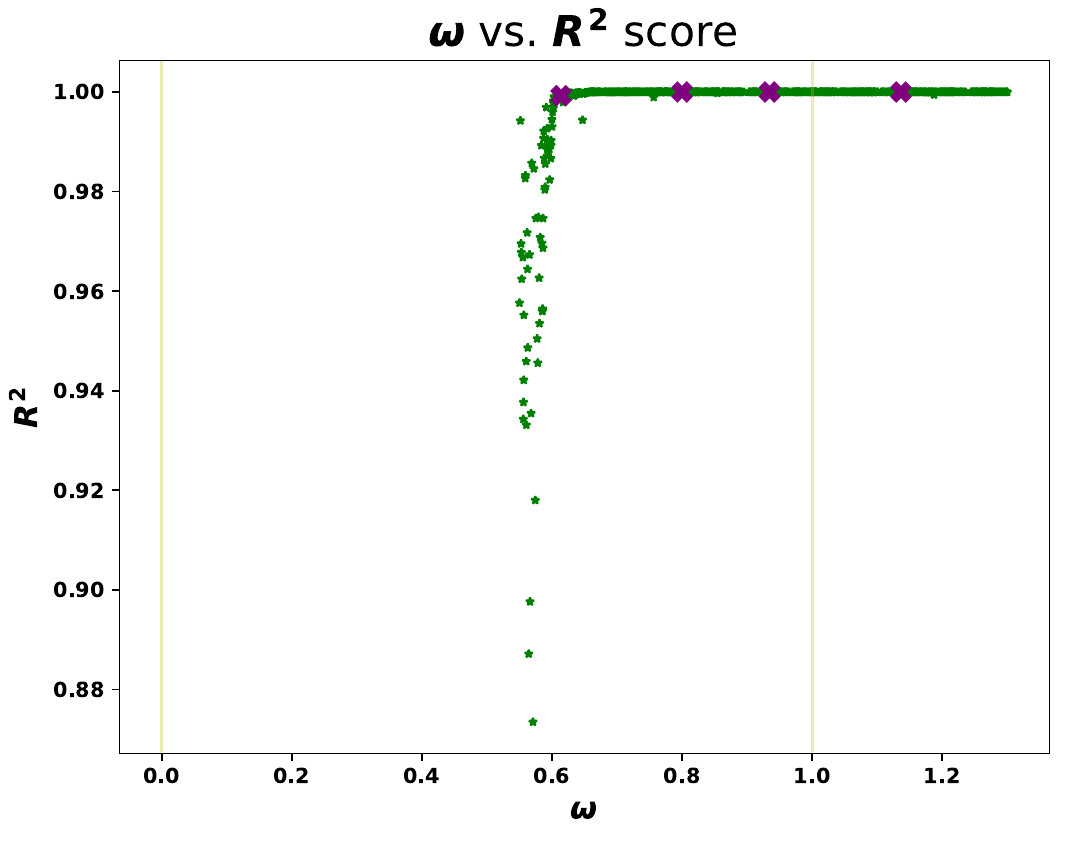}
    \caption{Predictive performance of the QRC across driving frequencies $\omega$ in the single-amplitude regime ($F_0 = 20$). Each point represents the $R^2$ score computed on the test set at a given frequency $\omega$, with an overall performance of $R^2 = 0.9997$.  }
    \label{fig:amp20}
\end{figure}

Five-fold cross-validation over the hyperparameter grid of Sec.~\ref{sec:problem_formulation} yields $\alpha = 10^{-4}$ and $\gamma_{\mathrm{rbf}} = 15.0$. The resulting narrow kernel is consistent with the structured, low-variance geometry of the PCA-compressed feature space along this single frequency axis. The trained readout yields $\mathrm{MSE} = 10^{-5}$ and
\begin{equation}
R^2 = 0.9997,
\end{equation}
reflecting near-perfect reconstruction of $\langle \hat{\sigma}_z \rangle(t)$ across the full test distribution. Notably, this level of accuracy is attained with a single reservoir qubit [$N=1$ in Eq.~\eqref{eq:LME}], a regime where neither entanglement nor exponential Hilbert-space scaling plays a role. The result implicates driven-dissipative dynamics as the operative computational resource.
Figure~\ref{fig:amp20} plots per-sample $R^2$ as a function of $\omega$ for all test realizations. Scores cluster tightly near unity across the full band; the sole exception is a small group of isolated dips just below resonance ($\omega \lesssim  = 1$). No significant degradation is observed at resonance itself, despite the strongly nonlinear character of the near-resonant response. This reflects the reservoir's capacity to encode instantaneously nonlinear, history-dependent features through its Lindblad-governed trajectory in Hilbert space.

The sub-resonant accuracy reduction is consistent with a memory-depth argument. As the system transitions from relaxation-dominated to drive-dominated dynamics with decreasing $\omega$, the effective temporal memory required to reproduce the target grows. The fixed dissipation rate $\gamma = 5.0$ sets a finite memory window that is well matched to most of the band but marginally insufficient for the longest-memory trajectories at the lowest frequencies probed. This is a tunable property of the reservoir rather than a fundamental constraint, and its effect remains small throughout.

From a representation standpoint, the $512$-dimensional reservoir feature vectors are compressible to a $15$-dimensional PCA subspace with no measurable accuracy loss. This dimensionality reduction indicates that the open-system dynamics populate a low-dimensional manifold within the full feature space, and the sharp RBF kernel ($\gamma_{\mathrm{rbf}} = 15.0$) is well adapted to its compact, low-dispersion geometry. 

\begin{figure}
    \centering
    \includegraphics[width=\columnwidth]{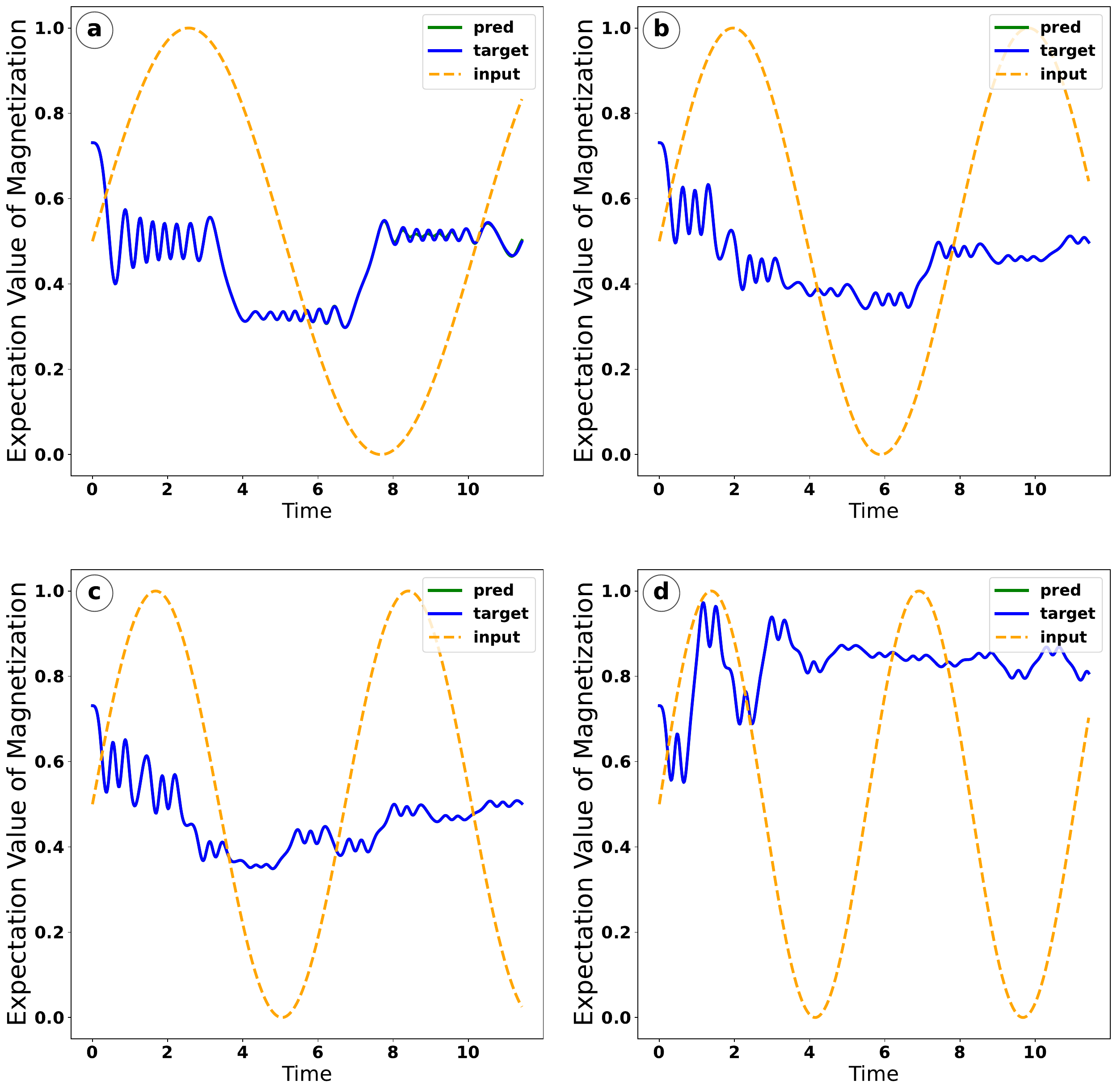}
    \caption{Performance illustration for the QRC model in the intermediate drive regime ($F_0 = 20$). Panels (a)--(d) show the input driving signal (input), the corresponding target time series (target), and the QRC-predicted output (pred) for driving frequencies $\omega \in \{0.61, 0.80, 0.93, 1.14\}$, respectively, highlighted by purple crosses in the $R^2$ vs $\omega$ plot. }
    \label{fig:amp20_pred}
\end{figure}

Figure~\ref{fig:amp20_pred} shows representative predicted and target time series at four frequencies $\omega \in \{0.61, 0.80, 0.93, 1.14\}$, spanning sub-resonant, near-resonant, and super-resonant regimes. In each panel the predicted trajectory faithfully traces the ground-truth observable $\langle \hat{\sigma}_z \rangle(t)$ across the full integration window, capturing both the slowly varying envelope and the fine oscillatory structure.

The data-efficiency tradeoff between reservoir size and training cost is summarized in Table~\ref{tab:qrc_performance}. For $N=1$, pass rates at both $R^2$ thresholds approach their maximum (99.59\% and 98.38\%, respectively) only at the full dataset size $N_\omega = 3700$. Augmenting the reservoir to $N=2$ qubits achieves $100\%$ at both thresholds at $N_\omega = 1000$, a reduction in training data of more than threefold with no modification to the readout architecture or training procedure. This accelerated convergence reflects the larger effective feature space of the two-qubit reservoir, which attains comparable representational coverage from substantially fewer input–output pairs. 
At first glance, the requirement of $N_\omega = 3700$ training trajectories for the single-qubit reservoir appears at odds with the characterization of the model as a compact surrogate. This reflects a distinction between architectural compactness and data efficiency. While the reservoir is minimal, its representational capacity must be populated through sampling of the input--output manifold. As shown by the $N=2$ results, modest increases in reservoir dimensionality substantially reduce the data requirement, indicating that the observed training cost is not intrinsic to the task but arises from the limited feature space of the minimal reservoir.

These results demonstrate that Lindblad-governed dynamics alone are sufficient to achieve near-perfect transduction in this setting. The near-perfect global $R^2$, uniform frequency coverage, and low-dimensional feature manifold collectively establish the sufficiency of driven-dissipative dynamics as a computational resource for this class of problem. These findings set the reference against which the reservoir's generalization capacity is assessed in Sec.\ref{sec:results_multi}, where the driving amplitude is varied over a $20\times$ range.

\begin{table}[t]
\centering
\caption{Performance of the DQRC readout as a function of the number of training data points $N_\omega$, for $N=1$ and $N=2$ reservoir qubits [ Eq.~\eqref{eq:LME}]. The $N=2$ case saturates at $100\%$ from $N_\omega=1000$ onward (entries omitted); $N=1$ reaches $R^2 > 0.90$ in $99.59\%$ and $R^2 > 0.95$ in $98.38\%$ of cases by $N_\omega = 3700$, indicating that adding a qubit accelerates convergence. }
\label{tab:qrc_performance}
\setlength{\tabcolsep}{8pt}
\begin{tabular}{ccccc}
\hline\hline
$N_\omega$ & $N$ & $R^2$ & $R^2 > 0.90$ (\%)& $R^2 > 0.95$ (\%)\\
\hline

1000 & 1 & 0.9984 & 94.5 & 92.0 \\
& 2 & 0.9999 & 100& 100\\
\hline

1500 & 1 & 0.9984 & 94.0 & 92.6 \\
\hline

2000 & 1 & 0.9989 & 95.0 & 94.0 \\
\hline

3700 & 1 & 0.9997 & 99.59& 98.38\\
\hline\hline

\end{tabular}
\end{table}

\subsection{Multi-amplitude dataset ($F_0 \in [1,\,20]$)}
\label{sec:results_multi}

The dissipation rate was held at $\gamma = 5.0$ across both the single- and multi-amplitude regimes, ensuring that the reservoir memory timescale remained constant and that any performance variation could be attributed unambiguously to the expanded input space rather than to changes in the reservoir itself. 

Having established near-perfect reconstruction in the single-amplitude regime, we impose the more stringent requirement of generalization across a $20\times$ variation in driving amplitude. This promotes the input space from a one-dimensional frequency grid to a two-dimensional $(\omega, F_0)$ manifold, spanning weakly driven near-linear dynamics at $F_0=1$ to strongly nonlinear behavior at  $F_0=20$. The amplitude coordinate is never explicitly appended to the feature vector; it enters exclusively through the reservoir dynamics, yet the readout resolves it directly from the resulting dynamical embedding. As quantified in Sec.~\ref{sec:complexity}, the amplitude-aware permutation entropy increases monotonically with $F_0$, establishing a controlled escalation in task complexity.

Five-fold cross-validation selects $\alpha = 5\times10^{-4}$ and
$\gamma_{\mathrm{rbf}} = 1.0$. Relative to the
single-amplitude optimum ($\alpha = 10^{-4}$, $\gamma_{\mathrm{rbf}} = 15.0$),
this shift toward a broader kernel reflects the increased dispersion of the feature distribution. This spread arose from variability in both frequency and amplitude. The sharper kernel in the single-amplitude case had been well-suited to a low-variance, structured feature space. In contrast, the present setting required a more flexible kernel to accommodate greater heterogeneity. The trained readout achieved
$\mathrm{MSE} = 3.2 \times 10^{-4}$ and a global coefficient of determination of
\begin{equation}
  R^{2} = 0.9709.
  \label{eq:r2_multi}
\end{equation}

The reduction from the single-amplitude value of $0.9997$ is moderate and expected, reflecting the transition to a two-dimensional parameter space together with the associated increase in signal complexity under a fixed reservoir and readout.

A per-sample breakdown is given in Table~\ref{tab:qrc_multi}.

\begin{table}[h]
\centering
\caption{Performance of the multi-amplitude QRC readout on the held-out
test set ($N_{\mathrm{test}} = 800$ samples, $N_{\mathrm{total}} = 4000$,
$F_0 \in [1,20]$), using a single QRC\,+PCA($d = 15$)+KRR model. Pass rates are reported at two $R^2$ thresholds.}
\label{tab:qrc_multi}
\setlength{\tabcolsep}{8pt}
\begin{tabular}{cccc}
\hline\hline
$N_{\mathrm{total}}$ & $R^2_{\mathrm{th}}$ & Passed / Total & Pass rate (\%) \\
\hline
$4000$ & $0.90$ & $710 / 800$ & $88.75$ \\
$4000$ & $0.95$ & $687 / 800$ & $85.88$ \\
\hline\hline
\end{tabular}
\end{table}

Table~\ref{tab:qrc_multi} shows that $88.75\%$ of test realizations exceed
$R^{2} > 0.90$ and $85.88\%$ exceed $R^{2} > 0.95$, indicating robust performance across the full amplitude range. From a representational standpoint, the amplitude coordinate effectively partitions the feature space into amplitude-stratified submanifolds, and the kernel readout interpolates smoothly across drive strengths within a single trained model. The residual performance gap relative to the single-amplitude regime tracks the growth of signal complexity with $F_0$, a connection formalized through the amplitude-aware permutation entropy analysis of Sec.~\ref{sec:complexity}. This indicates that the observed behavior is governed by the informational complexity of the target signals rather than by any structural limitation of the reservoir.

\begin{figure}
    \centering
    \includegraphics[width=\columnwidth]{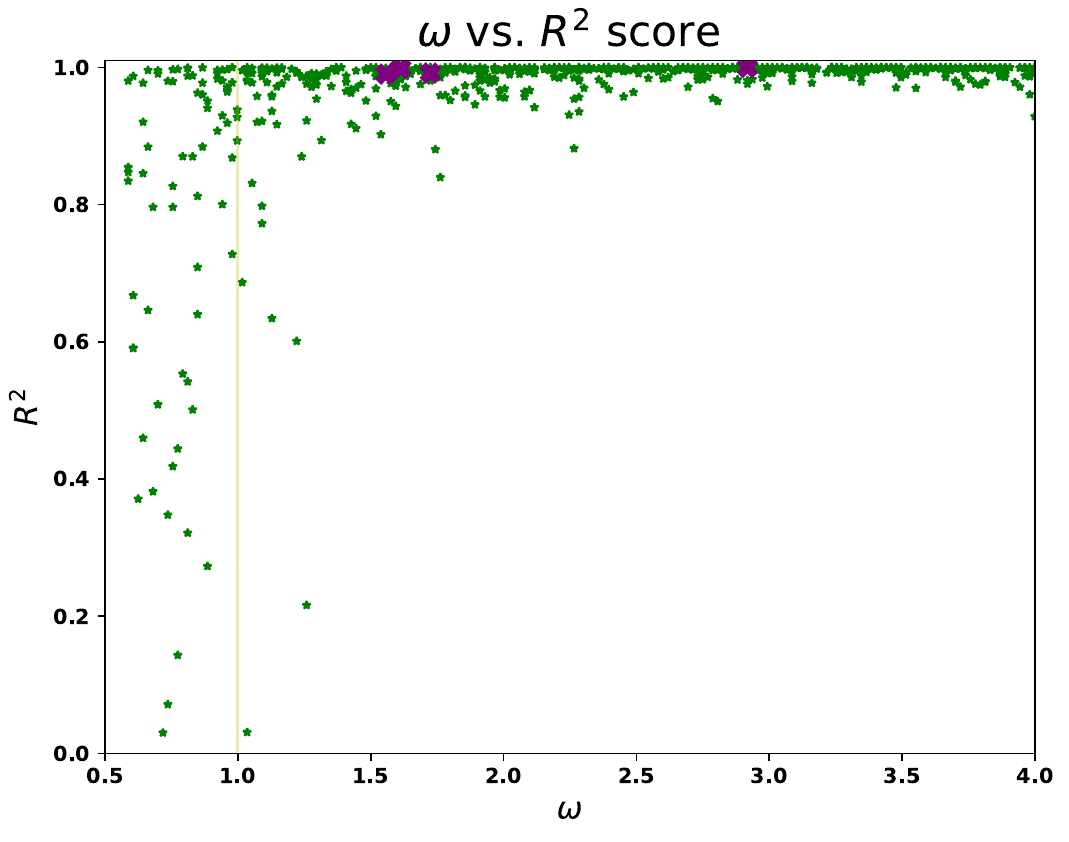}
    \caption{Predictive performance of the QRC across driving frequencies $\omega$ in the multi-amplitude regime ($F_0  \in [1,\,20]$). Each point represents the $R^2$ score computed on the test set at a given frequency $\omega$, with an overall performance of $R^2 = 0.9709$. }
    \label{fig:r2_omega_multi}
\end{figure}

Figure~\ref{fig:r2_omega_multi} shows the per-sample $R^{2}$ scores as a
function of $\omega$ for all $800$ test realizations across $F_0  \in [1,\,20]$. The plot reveals two patterns: $R^2$ scores are systematically higher at intermediate and super-resonant frequencies, where the response is more regular, and, at fixed frequency, they decrease with increasing $F_0$, tracking the monotonic growth of permutation entropy documented in Sec.~\ref{sec:complexity}. The sharpest degradation appears near resonance at high amplitude, where the signal complexity is highest. Even there, the predictions remain well behaved, suggesting a stabilizing role of the dissipative dynamics.

\begin{figure}
    \centering
    \includegraphics[width=\columnwidth]{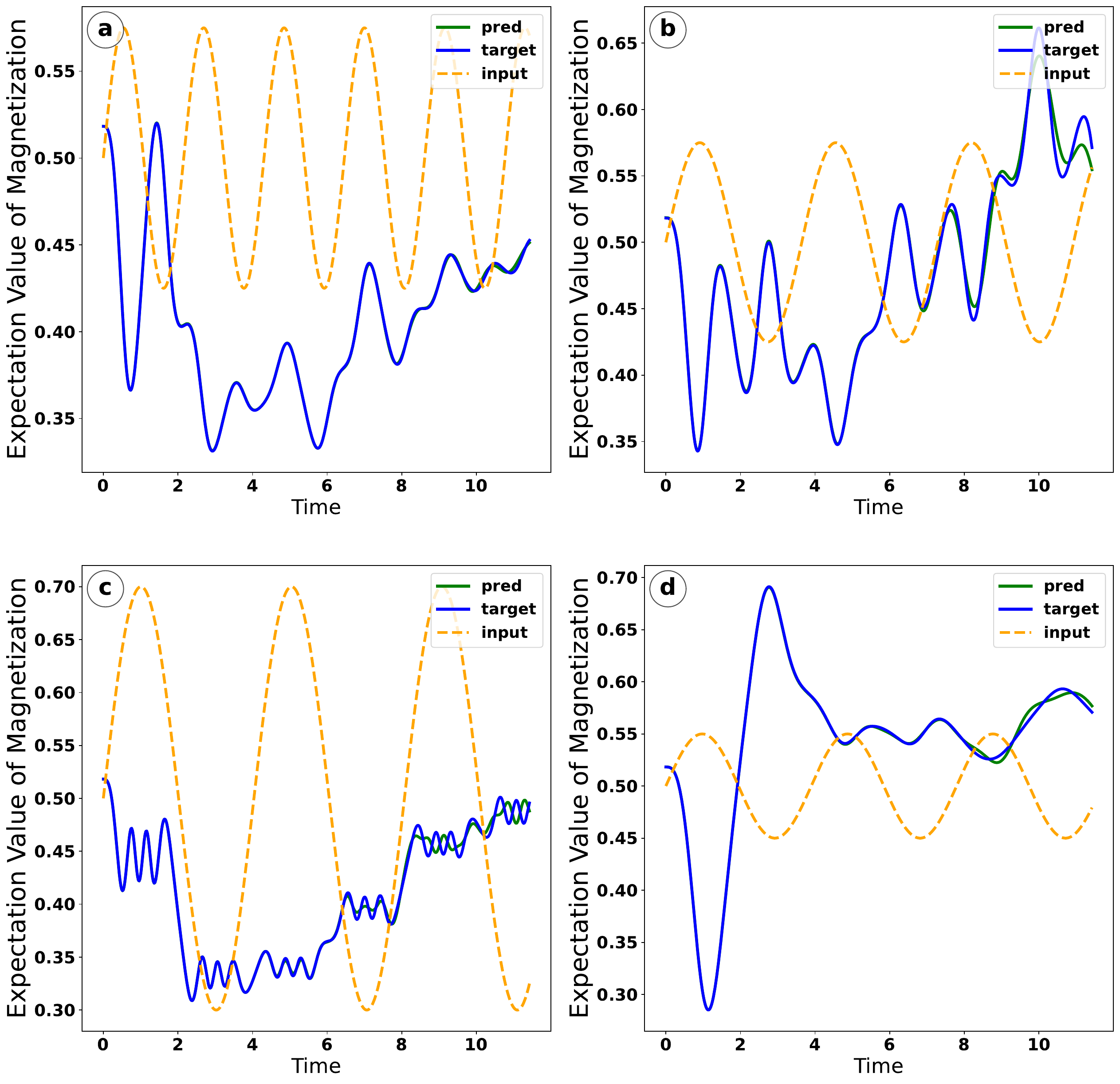}
    \caption{Performance illustration for the QRC model in the multi-amplitude regime ($F_0 \in [1,\,20]$). Panels (a)--(d) show the input driving signal (input), the corresponding target time series (target), and the QRC-predicted output (pred) for driving frequencies $\omega \in \{2.92, 1.72, 1.55, 1.62\}$, respectively, highlighted by purple crosses in the $R^2$ vs $\omega$ plot. }
    \label{fig:pred_multi}
\end{figure}

Figure~\ref{fig:pred_multi} shows representative predictions for test indices
$\{2,\,100,\,250,\,600\}$, spanning a range of driving frequencies and amplitudes. In each panel, the predicted trajectory (green) is overlaid against the
ground-truth observable (blue) and the
normalized input drive (orange dashed) over the full integration
window. The readout accurately traces both the slowly varying envelope and the fine oscillatory structure in most cases. Residual errors, when present, appear as phase drift or partial amplitude mismatch and are largely confined to the low-frequency, high-amplitude regime of highest complexity. The connection between this accuracy profile and the underlying signal complexity is taken up in Sec.~\ref{sec:complexity}.

\subsection{Output Complexity Analysis}
\label{sec:complexity}

To quantify the informational structure of the target signals across the full
parameter space $(\omega, F_0)$, we compute the amplitude-aware
permutation entropy, originally introduced by Bandt and Pompe~\cite{Bandt2002} and later extended to incorporate amplitude information by Fadlallah \emph{et al.} \cite{Fadlallah2013},
\begin{equation}
    H_{\mathrm{amp}}(m) = -\sum_{\pi} \tilde{p}(\pi)\, \log \tilde{p}(\pi),
    \label{eq:APE}
\end{equation}
where the amplitude-weighted pattern probability is defined as
\[
\tilde{p}(\pi) = \frac{\sum_{t:\,\Pi(t)=\pi} w(t)}{\sum_t w(t)},
\]
with
\[
w(t) = \frac{1}{m}\sum_{j=0}^{m-1} \lvert x(t+j) \rvert
\]
the mean absolute amplitude of the subsequence beginning at time $t$, and $\Pi(t)$ the ordinal pattern associated with that subsequence~\cite{Fadlallah2013}. Compared with standard permutation entropy, this
formulation assigns greater weight to high-variance segments, jointly capturing
ordinal structure and amplitude information. We use embedding dimension $m = 4$,
unit delay $\tau = 1$, and normalize by $\log(m!)$ so that $H_{\mathrm{amp}}
\in [0,1]$. For each amplitude $F_0$, the reported value is obtained by
averaging $H_{\mathrm{amp}}$ over all driving frequencies $\omega$.

\begin{figure}
    \centering
    \includegraphics[width=\columnwidth]{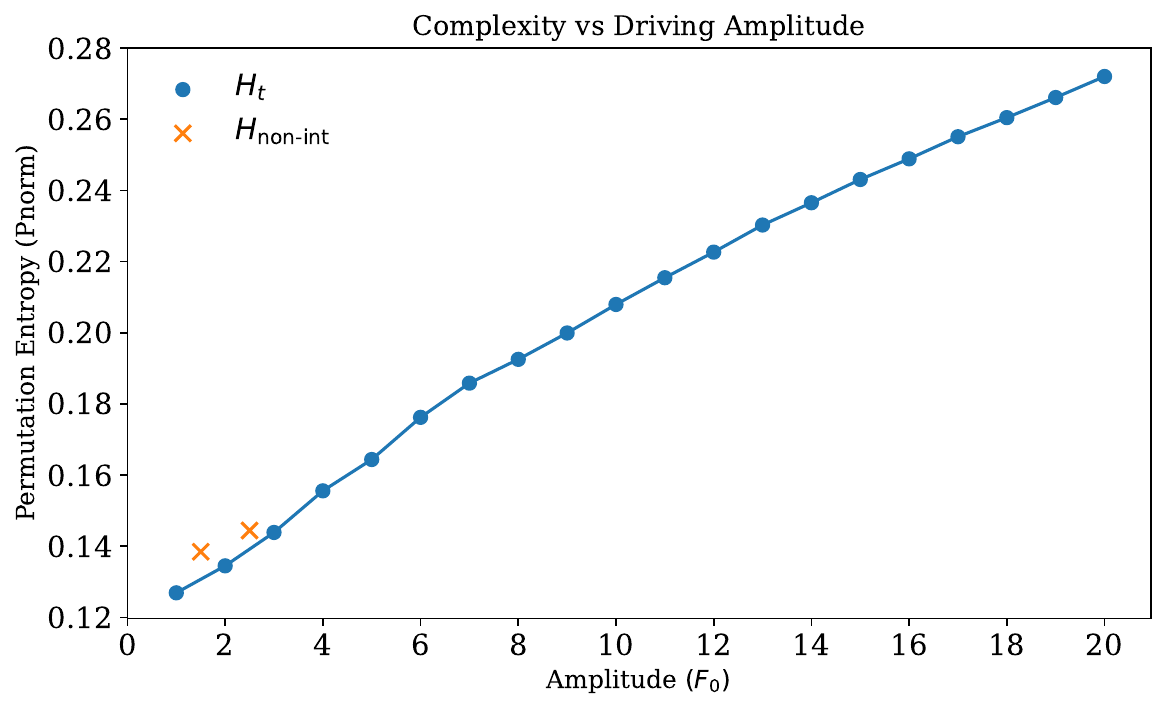}
    \caption{Amplitude dependence of the averaged amplitude-aware permutation entropy. The increasing trend with $F_0$ indicates progressive enrichment of nonlinear dynamics at higher drive strengths.  }
    \label{fig:complexity}
\end{figure}

Figure~\ref{fig:complexity} shows that the complexity exhibits an overall
increasing trend with $F_0 \in [1,20]$: at low amplitudes the response is nearly
linear and low-dimensional, whereas larger drives induce stronger nonlinear
mixing and richer temporal structure.

This entropy gradient tracks the $R^2$ degradation seen in the multi-amplitude results. In the single-amplitude case ($F_0 = 20$), the readout
operates over a relatively narrow distribution of target signals, yielding near-perfect accuracy ($R^2 \approx 1$). The multi-amplitude case spans the full
two-dimensional $(\omega, F_0)$ space, where $H_{\mathrm{amp}}$ is both broader
and shifted toward higher values, consistent with the reduction in global $R^2$
to $0.9709$. The systematic co-variation of per-sample $R^2$ scores with
$H_{\mathrm{amp}}$, discussed in Sec.~\ref{sec:results_multi}, indicates that the
observed accuracy profile reflects dataset complexity rather than a structural
limitation of the reservoir.
As a consistency check, we note that within the single-amplitude regime ($F_0 = 20$), where $H_{\mathrm{amp}}$ varies only with frequency, the per-sample $R^2$ scores in Fig.~\ref{fig:amp20} exhibit their largest deviations in the sub-resonant band where the permutation entropy is highest. This within-amplitude correspondence, where entropy varies only with frequency while $F_0$ is fixed, isolates the role of signal complexity from drive strength, providing further evidence that $H_{\mathrm{amp}}$ captures prediction difficulty independently of amplitude.

\subsection{Discussion}
\label{sec:discussion}

The results from both amplitude regimes demonstrate that a single fixed reservoir generalizes across a $20\times$ variation in drive strength without explicit feature augmentation, as amplitude information is implicitly encoded in the reservoir dynamics. The near-perfect accuracy in the single-amplitude regime ($R^2 = 0.9997$) establishes that the quantum reservoir accurately reconstructs the frequency-dependent response of the driven spin chain from input-output data. The reduction to $R^2 = 0.9709$ in the multi-amplitude regime
is consistent with the complexity analysis of Sec.~\ref{sec:complexity}:
the co-variation of per-sample $R^2$ scores with $H_{\mathrm{amp}}$ across both
frequency and amplitude suggests that residual errors are governed by the
intrinsic informational content of the target signals rather than by model
capacity.

The implicit amplitude conditioning via reservoir dynamics, which requires no explicit feature augmentation, accounts for much of this robustness. The contractive Lindblad damping further stabilizes the dynamics, keeping predictions physically bounded even in the hardest corner of parameter space.

The primary bottleneck identified in this study arises from the elevated signal complexity in the high-amplitude, near-resonance regime, where $H_{\mathrm{amp}}$ is largest and $R^2$ scores are most dispersed. Amplitude-stratified training or nonlinear readout layers appear to be promising approaches to close this gap, and we leave a systematic comparison to future work. Beyond this specific problem, amplitude-aware permutation entropy may serve as a useful pre-hoc diagnostic for estimating task difficulty before committing to a reservoir architecture, a point that merits further exploration in classical RC as well. 

\section{Conclusion}
\label{sec:conclusion}

We have introduced a dissipative quantum reservoir computing framework that acts as a digital twin of a driven many-body quantum system, learning its input--output map directly from trajectories while keeping the reservoir fixed and training only a lightweight classical readout. The surrogate reproduces the nonlinear, history-dependent response with high fidelity: $R^2 = 0.9997$ in the single-amplitude regime and $R^2 = 0.9709$ across a $20\times$ variation in drive strength, with amplitude resolved implicitly from the dissipative dynamics rather than appended as a feature. Most striking is that this is delivered by a trivially small reservoir:  the single-qubit ($N=1$ in Eq.~\eqref{eq:LME}) reservoir matches and on several metrics surpasses the physics-informed Kolmogorov--Arnold-network model of Ref.~\cite{Sen2025KAN} and improves decisively on the black-box temporal convolutional approach of Ref.~\cite{Sen2025TCN}, with a single- or two-qubit reservoir faithfully reconstructing the observable dynamics of a ten-qubit Ising chain.

We interpret this compression as an automatic, data-driven discovery of an effective quasiparticle description~\cite{mattuck2012guide}. The quasiparticle is arguably the single most fundamental concept of condensed matter physics: the low-energy behavior of a strongly interacting many-body system is almost never described by its bare microscopic degrees of freedom, but by a small number of collective noninteracting excitations. Our results recover precisely this picture. The response of the $N=10$ chain, formally living in a $2^{10}$-dimensional Hilbert space, is carried by only a few emergent modes, and the driven-dissipative reservoir plays the role of the quasiparticle, a single effective degree of freedom that suffices to reproduce the many-body observable. This viewpoint clarifies why entanglement and raw Hilbert-space dimension appear largely irrelevant here: the operative resource is the temporal memory-limiting action of the open-system dynamics, which lets a minimal reservoir stand in for a much larger medium by reproducing its emergent, rather than its microscopic, structure. The search for quantum advantage in temporal learning may thus be better guided by physical faithfulness~\cite{Ehrenfest1927, Bondar2012} and engineered dissipation~\cite{Sannia2024,Fujii2017,Nakajima2019} than by Hilbert-space scaling alone.

\begin{acknowledgments}
This work was supported by Army Research Office (ARO) (grant W911NF-23-1-0288; program manager Dr.~James Joseph). The views and conclusions contained in this document are those of the authors and should not be interpreted as representing the official policies, either expressed or implied, of ARO, or the U.S. Government. The U.S. Government is authorized to reproduce and distribute reprints for Government purposes notwithstanding any copyright notation herein.
\end{acknowledgments}

\section*{Data availability}
All codes used in this study can be found in \href{https://github.com/AI-and-Quantum-Computing/DQuRC}{https://github.com/AI-and-Quantum-Computing/DQuRC}.

% ============================ APPENDICES ==============================
%\appendix
%\section{mmmmmm}

% ============================ REFERENCES =============================
\bibliography{references}

\end{document}